\documentstyle[prd,aps,floats,graphicx]{revtex}  %% RevTeX V 3.1
\def\be{\begin{equation}}\def\ee{\end{equation}}
\def\bea{\begin{eqnarray}}\def\eea{\end{eqnarray}}
\def\goesas{\mathop{\sim}\limits} \def\LA{\Lambda} \def\etal{{\it et al.}}
 \def\rh{\rho} \def\ph{\phi} \def\qn{q_0} 
\def\Om{\Omega} \def\gm{\gamma} \def\kp{\kappa} \def\rarr{\rightarrow}
 \def\Omp{\Om_{\ph0}} \def\la{\lambda}
\begin{document}
\draft
\twocolumn[\hsize\textwidth\columnwidth\hsize\csname
@twocolumnfalse\endcsname
\title{Observational constraints upon quintessence models arising from moduli 
fields}
\author{S.C. Cindy Ng\cite{Ecng}}
\address{Department of Physics and Mathematical Physics, University of
Adelaide,\break Adelaide, S.A. 5005,~~~Australia.}
\date{2 Jun, 2000; ADP-00-19/M92, astro-ph/0004196}
\maketitle

\begin{abstract}
We study observational constraints on cosmological models with a quintessence
arising from moduli fields. The scalar field potential is given by a double
exponential potential $V=V_0\exp(-Ae^{\sqrt{2}\kp\ph})$. After reviewing the 
properties of the solutions, from a dynamical systems phase space analysis, we 
consider the constraints on parameter values imposed by luminosity distances 
from the 60 Type IA supernovae published by Perlmutter \etal, and also from 
gravitational lensing statistics of distant quasars. We also update the 
constraints on models with a single exponential potential 
$V=V_0e^{-\la\kp\ph}$. 
\end{abstract}
\pacs{PACS numbers: 98.80.Cq 95.35.+d 98.80.Es}
\vskip2pc]

\section{Introduction}

A cosmological constant $\LA$ has been considered as the missing energy of the 
universe for a long time. Recently, a varying vacuum energy or 
``quintessence'' \cite{CDS} has become popular as an alternative candidate.
Since more parameters are involved in such models, they are more flexible in 
explaining some of the problems left by the cosmological constant model. One 
such problems is the ``cosmic coincidence problem'' \cite{ccp}: the missing 
energy and the matter energy densities decrease at different rates as the 
universe expands, so it seems purely coincident that they are comparable today 
with the concordant values from several observational tests of 
$\Omp\goesas{2\over3}$ and $\Om_{m0}\goesas{1\over3}$ \cite{WCOS}. 
Several candidates for quintessence fields have been proposed. Typically
quintessence models possess attractor solutions with common evolutionary 
properties for a wide range of initial conditions. For example, in some cases 
the scalar field energy maintains constant ratio to the matter density at late
time \cite{PNGB}, where as in other models it tracks the dominant matter
component in some other sense \cite{plp,ss,CLW,FJ,LS}.

In this paper, I will study quintessences arising from string moduli. In 
string or Kaluza-Klein type models the moduli fields associated with the 
geometry of the extra dimensions may have effective potentials which depend
exponentially on the moduli fields, due to the curvature of the internal 
spaces, or alternatively through the interaction of the moduli with the form 
fields on the internal spaces (see, e.g., \cite{CLW} and references therein). 
Single exponential potentials of the form
\be \label{sep} V=V_0e^{-\la\kp\ph} \ee
which give rise to a scaling solution have been well-studied in the 
literatures (see, e.g., \cite{ss,CLW,FJ} and references therein). In this 
paper I will concentrate on double expontial potentials of the form
\be \label{dep} V=V_0\exp(-Ae^{\sqrt{2}\kp\ph})\,  \ee
which typically arise from supersymmetry breaking via gaugino condensation 
\cite{gc,BCC,Bin}. Furthermore, it has also recently been argued that if the 
superpotential of the moduli field in string theory is T-duality invariant, 
then it cannot be approximated by a single exponential function, but must 
depend on double exponentials \cite{delaM}.

Binetruy \cite{Bin}, and later de la Macorra \cite{delaM}, have argued that 
scalar fields with double exponential potentials cannot act as quintessences
for two reasons. Firstly, such models have a global attractor solution leading 
to a matter-dominated universe at late times. Secondly, near the attractor the 
equation of state is positive, $w_\ph>0$ and approaching zero as 
$t\rarr\infty$. This would appear to contradict the latest observational 
results from type IA supernovae that a quintessence is dominating over matter 
and the universe is entering a phase of accelerated expansion \cite{Per,Rie}, 
similar to inflation. The purpose of this paper is to show that for parameter
values away from the attractor, there can exist models which are consistent 
with the observational tests, although some tuning is required for the 
universe not to have reached the attractor at present. This paper also updates 
the observational constraints on the single exponential potential model, which
have been given by Frieman and Waga \cite{FW}.

The organization of this paper is as followed: in section II, I will present a 
phase-space analysis of the double exponential potential model. In section 
III, I will discuss the numerical integration of the evolution equations for 
both the single and double exponential potential models, and obtain values for 
$\Omp$ and $H_0t_0$. In section IV, I will constrain both models using the 
light-curve calibration luminosity distances of type IA supernovae (see 
\cite{Per,Dre} and references therein) and the gravitational lensing 
statistics of high luminosity quasars by intervening galaxies (see 
\cite{Koc,WM} and references therein). Similar constraints on other 
quintessence models have been presented in \cite{FW,WM,WF,NW}.

\section{Phase space analysis for double exponential potential model}

We begin by considering a universe which consists of a scalar field with the 
potential (\ref{dep}) and a barotropic fluid with equation of state 
$P_\gm=(\gm-1)\rh_\gm$, $0\leq\gm\leq2$. For a spatially-flat 
Friedmann-Robertson-Walker (FRW) universe, the governing equations are given by
\bea
\label{Hdot}\dot{H}&=&-{\kp^2\over2}\left(\rh_\gm+P_\gm+\dot{\ph}^2\right)\ , 
\\ \label{rhgmdot}\dot{\rh_\gm}&=&-3H(\rh_\gm+P_\gm )\ , \\ 
\label{phdot}\ddot{\ph}&=&-3H\dot{\ph}-{dV\over d\ph}\ , 
\eea
subject to the Friedmann constraint
\be
\label{fc}H^2={\kp^2\over3}\left(\rh_\gm+{1\over2}\dot{\ph}^2+V\right)\ ,
\ee
where an overdot denotes the ordinary differentiation with respect to the time 
$t$. In the above equations, $\kp^2\equiv8\pi G$, and $H\equiv\dot{a}/a$ is 
the Hubble parameter.

We follow the treatment of Copeland, Liddle and Wands \cite{CLW} by defining 
the variables 
\be \label{xy}x=\frac{\kp\dot{\ph}}{\sqrt{6}H}\ ,
y=\frac{\kp\sqrt{V}}{\sqrt{3}H}\ , \ee
and introducing a third variable \cite{MP}
\be \la(\ph)=-{dV/d\ph\over\kp V}=\sqrt{2}Ae^{\sqrt{2}\kp\ph}\ . \ee
In the case of the single exponential potential (\ref{sep}), $\la(\ph)$ 
reduces to a constant. In terms of these variables the evolution equations 
(\ref{Hdot}) -- (\ref{fc}) become 
\bea
\label{xprime}x'&=&-3x+\la\sqrt{3\over2}y^2+{3\over2}x\left[2x^2+
\gm(1-x^2-y^2)\right]\ , \\ \label{yprime}y'&=&-\la\sqrt{3\over2}xy+
{3\over2}y\left[2x^2+\gm(1-x^2-y^2)\right]\ , \\
\label{laprime}\la'&=&2\sqrt{3}x\la\ ,
\eea
where a prime denotes a derivative with respect to the logarithm of the scale 
factor, $N\equiv\ln(a)$. The system is bounded by the cylinder $x^2+y^2=1$. 
Assuming that $A$ is positive, the system is confined to $\la\ge0$. The system
is symmetric under the reflection $y\rarr-y$.

\begin{figure}[ht] \includegraphics[height=6cm,width=8cm]{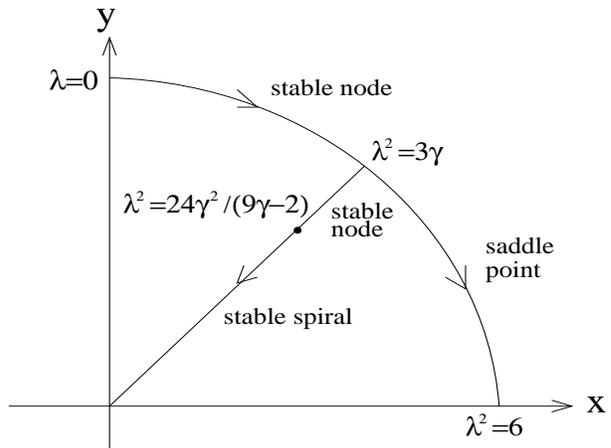}
\caption{\label{attractor} The evolution of the late times attractors of the 
single exponential potential model as $\la^2$ increases. The scalar field 
dominated solution has $x=\la/\sqrt{6}$, $y=\sqrt{1-\la^2/6}$; the scaling 
solution has $x=\sqrt{3/2}\gm/\la$, $y=\sqrt{3(2-\gm)\gm/2\la^2}$. The scalar 
field dominated solution is a saddle point for $\la^2>3\gm$.} \end{figure}

As many properties of the solutions in the double exponential potential model 
can be related to the single exponential potential model, let us review the 
single exponential potential model as has been studied in the literature 
(\cite{ss,CLW,FJ} and references therein). Depending on the values $\la$ and 
$\gm$, there are up to five critical points in the $x$ -- $y$ plane, two of 
which are related to late times. For $\la^2<3\gm$, the late times attractor is 
a scalar field dominated solution $\Om_\ph\equiv\kp^2\rh_\ph/3H^2\rarr1$ with 
$\gm_\ph\equiv(\rh_\ph+P_\ph)/\rh_\ph\rarr{\la^2\over3}<\gm$. For 
$\la^2>3\gm$, the late times attractor is the scaling solution \cite{ss} with
$\gm_\ph=\gm$ and $\Om_\ph\rarr{3\gm\over\la^2}$, a constant. 
Fig.~\ref{attractor} shows the late times attractors and how they evolve as
$\la^2$ increases from zero. As $\la^2$ increases the attractor evolves from a
stable node into a stable spiral at $\la^2={24\gm^2\over(9\gm-2)}$. The 
attractor approaches the origin as $\la^2\rarr\infty$. 

In the double exponential potential model, we can identity critical points at 
finite $\la$ from the evolution equations (\ref{xprime}) -- (\ref{laprime}).
They are three discrete points at $(x_c,y_c,\la_c)=$ $(\pm1,0,0)$, $(0,1,0)$,
and a 1-parameter family at $(0,0,\la_c)$ for $\la_c$ arbitrary. The stability
of the critical points can be obtained by linearizing (\ref{xprime}) -- 
(\ref{laprime}) about the these points and solving for the eigenvalues of 
small perturbations. The critical points and the eigenvalues are listed in 
Table \ref{cpt}. 

\begin{table}[ht] \centering \caption[cpt]{\label{cpt} The critical points
at finite $\la$ and their eigenvalues for the double exponential potential 
model.} \bigskip 
\begin{tabular}{l|l|l|l} \noalign{\smallskip} $x_c$ & $y_c$ & $\la_c$ & 
Eigenvalues \\ \noalign{\smallskip} \hline \noalign{\smallskip}
$1$ & $0$ & $0$ & $3$, $3$, $2\sqrt{3}$ \\ 
$-1$ & $0$ & $0$ & $3$, $3$, $-2\sqrt{3}$ \\ 
$0$ & $1$ & $0$ & $-3$, $-3$, $0$\\
$0$ & $0$ & $0\le\la_c<\infty$ & $-3/2$, $3/2$, $0$ \\
\noalign{\smallskip} \end{tabular} \end{table}

The scalar field kinetic energy dominated solution $(1,0,0)$ is an unstable 
node, while $(-1,0,0)$ is a saddle point since trajectories are repelled in 
the $x$ and $y$ directions and are attracted in the $\la$ direction. The 
scalar field dominated solution $(0,1,0)$ attracts a two-dimensional bunch of 
trajectories but is degenerate in the $\la$ direction. For the barotropic 
fluid dominated solution $(0,0,\la_c),0\le\la_c<\infty$, trajectories are
attracted in the $x$ direction, repelled in the $y$ direction and degenerate
in the $\la$ direction. 

To examine the critical points at infinite $\la$, we use the 
Poincar\'{e} sphere method by transforming 
\be
\la={1\over\epsilon}\ .
\ee
On the surface $\epsilon=0$ we find
\bea
{dx\over d\tau}=\sqrt{3\over2}y^2\, \\
{dy\over d\tau}=-\sqrt{3\over2}xy\, 
\eea
where $\tau$ is a new time coordinate defined by $d\tau=\la dN$. Therefore, 
there are infinitely many critical points lying on the $\epsilon=0$ surface 
with $y_c=0$ and arbitrary $x_c\in[-1,1]$. The eigenvalues for the matrix of
perturbations are $0$, $-\sqrt{3\over2}x_c$, $0$. Therefore the critical 
points are degenerate in both the $x$ and $\la$ direction. They repel or 
attract trajectories in the $y$ direction depending on $-1\le x_c<0$ or 
$0< x_c\le1$ respectively. For $x_c=0$ all the eigenvalues vanish, and 
higher-order perturbations show that it is a stable attractor. 

\begin{figure}[ht] \includegraphics[height=6cm,width=8cm]{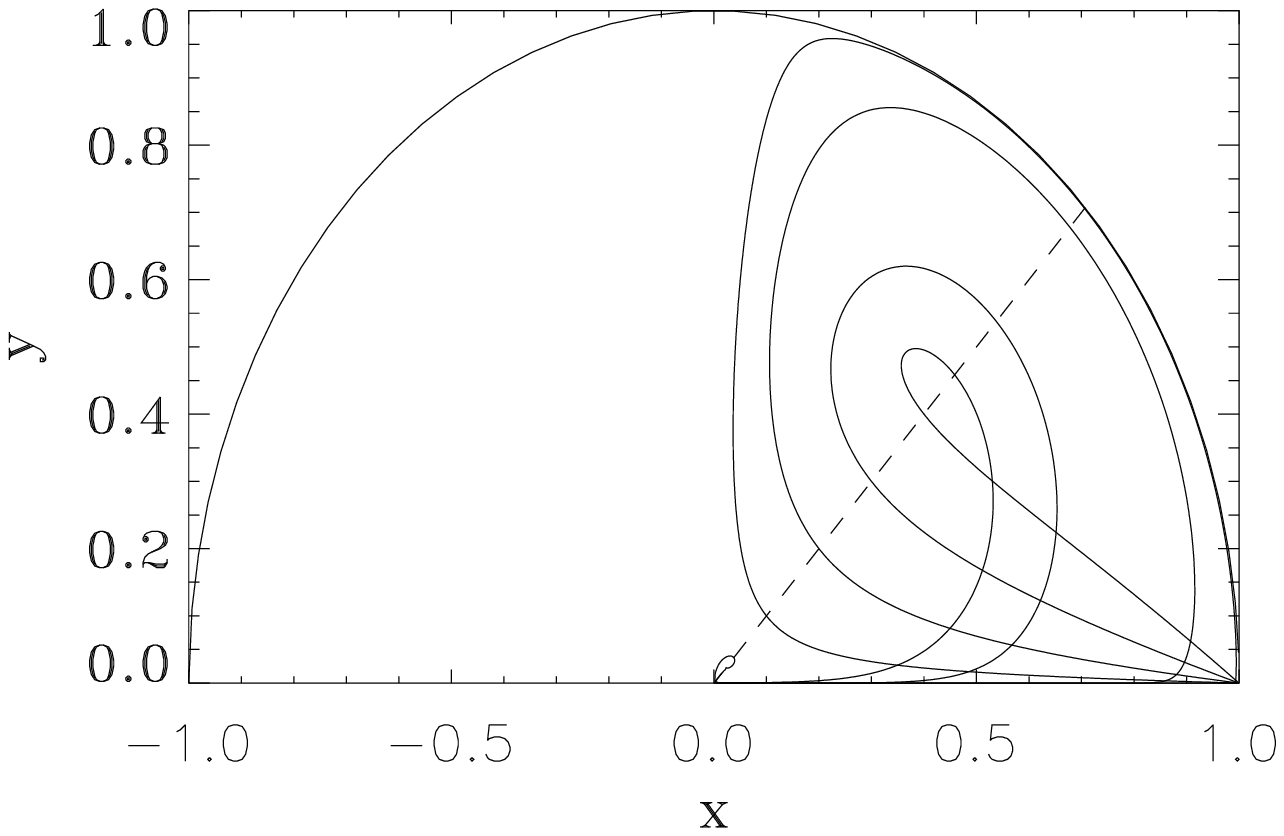}
\caption{\label{xy1} The projection of the space trajectories on the $x$ -- 
$y$ plane for $\gm=1$. The trajectories are for initial values 
$(x_i,y_i,\la_i)$: $(0.1,0.1,0.3)$, $(0.2,0.2,0.5)$, $(0.3,0.3,1.0)$, 
$(0.4,0.4,1.5)$. The dashed line is $x=y$.} 
\includegraphics[height=6cm,width=8cm]{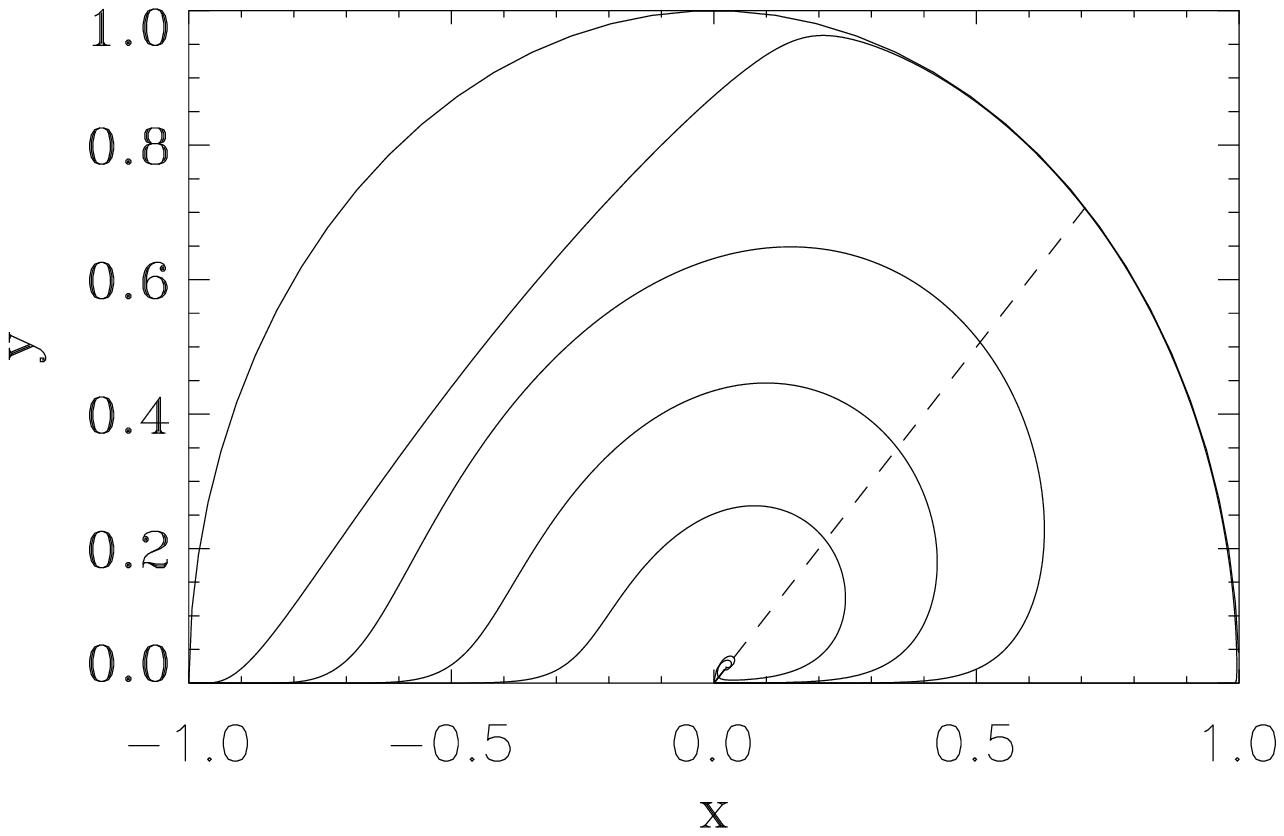} 
\caption{\label{xy2} The projection of the space trajectories on the $x$ 
-- $y$ plane for $\gm=1$. The trajectories are for the initial values 
$(x_i,y_i,\la_i)$: $(-0.1,0.2,15)$, $(-0.1,0.4,10)$, $(-0.1,0.6,5)$, 
$(-0.1,0.8,0.5)$. The dashed line is $x=y$.} \end{figure}

We will numerically integrate the evolution equations (\ref{xprime}) --
(\ref{laprime}) forwards and backwards from some initial point 
$(x_i,y_i,\la_i)$, in the case of a matter-dominated universe $\gm=1$. 
In Fig.~\ref{xy1} and \ref{xy2}, we project the trajectories onto the $x$ -- 
$y$ plane. When integrating backwords from the initial points 
$(x_i,y_i,\la_i)$, the trajectories in Fig.~\ref{xy1} approach the critical 
point $(1,0,0)$, and the trajectories in Fig.~\ref{xy2} approach the critical 
points $(x_c,0,\la_c)$, $x_c\in[-1,0)$, $\la\rarr\infty$.  

The scalar field is rolling down the potential at sufficiently late times so 
$\ph$ and hence $\la(\ph)$ is increasing with time. As $\la(\ph)$ increases 
with time, the behaviour of the solutions effectively approximates that of 
single exponential potential models, but with the parameter $\la$ evolving in 
the direction shown in Fig.~\ref{attractor}. For sufficiently small $\la_i$ 
the scalar field-dominated solution is dynamically important and the 
trajectories get very close to the circular boundary. As $\la$ increases, the 
solutions approach scaling solutions ($x=y$ for $\gm=1$) before spiralling 
towards the origin.

\section{Numerical integration}

In order to determine $\Omp$ and $H_0t_0$, and the luminosity distance 
-- redshift relation in the next section, we ought to integrate the evolution 
equations (\ref{Hdot}) -- (\ref{fc}) numerically. In this section, we will 
discuss the numerical integration for both the single and double exponential 
potential models in a manner similar to a previous study on the model with a 
pseudo Nambu-Goldstone boson (PNGB) potential \cite{NW}.

\subsection{Single exponential potential}

Here we update the numerical integration result for the single exponential 
potential by Frieman and Waga \cite{FW}. We use an alternate set of 
dimensionless variables: 
\be
\label{uvw}u={\kp\dot{\ph}\over H_0}\ , 
v=\Om_{m0}^{1/3}(1+z)\ , 
w=\la\kp\ph-\ln{\kp^2V_0\over H_0^2}\ ,
\ee
the variables $u$ and $v$ are the same as were used in \cite{NW}. In terms of 
these variables, the field equations become:
\bea 
\label{uprime}u'&=&-3{H\over H_0}u+\la e^{-w}\ , \\
v'&=&-{H\over H_0}v\ , \\
w'&=&\la u\ ,
\eea
where a prime denotes a derivative with respect to the dimensionless time 
parameter $H_0t$. The Hubble parameter is defined implicitly according to
\be \label{Hnorm}
{H\over H_0}\equiv\left(v^3+{1\over6}u^2+{1\over3}e^{-w}\right)^{1\over2}\ .
\ee

We shall begin the integration at last-scattering. We choose $H_0t_i=0$. At 
this initial stage, $H/H_0\gg1$ and the field evolution (\ref{uprime}) is 
overdamped by the expansion of the universe, driving $u$ to zero. We therefore 
choose $u_i=0$. At last-scattering, $z\goesas1100$ and 
$\Om_{m0}^{1/3}\goesas1$ for $\Om_{m0}\goesas1$, hence $v_i=1100$ is a 
reasonable value. Note that results of the integration do not change 
significantly if $H_0t_i$, $u_i$ and $v_i$ are altered to values within the 
same order of magnitude. The remaining parameters are $w_i$ and $\la$, which 
are determined by the physical origin of the model.

\begin{figure}[ht] \includegraphics[height=7cm,width=8cm]{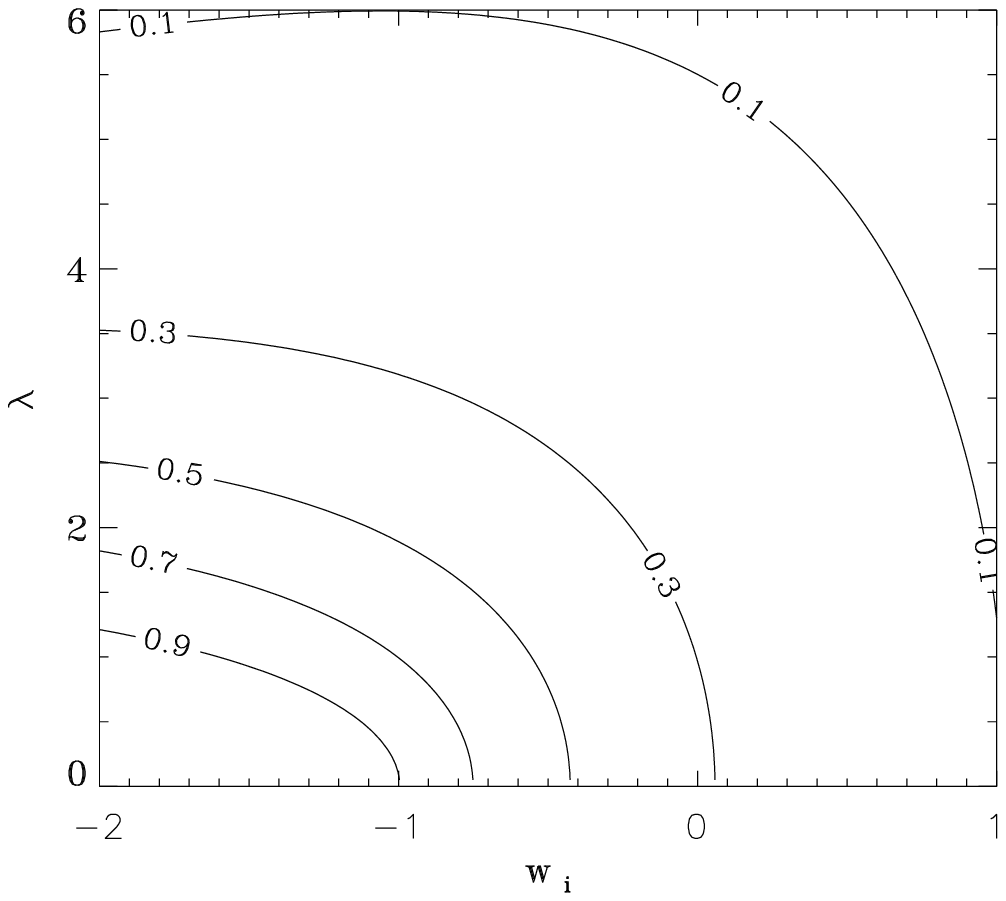}
\caption{\label{expV-Omph0} Contours of constant $\Omp$ in the $w_i$ -- $\la$ 
plane for the single exponential potential model.} \end{figure}

\begin{figure}[ht] \includegraphics[height=7cm,width=8cm]{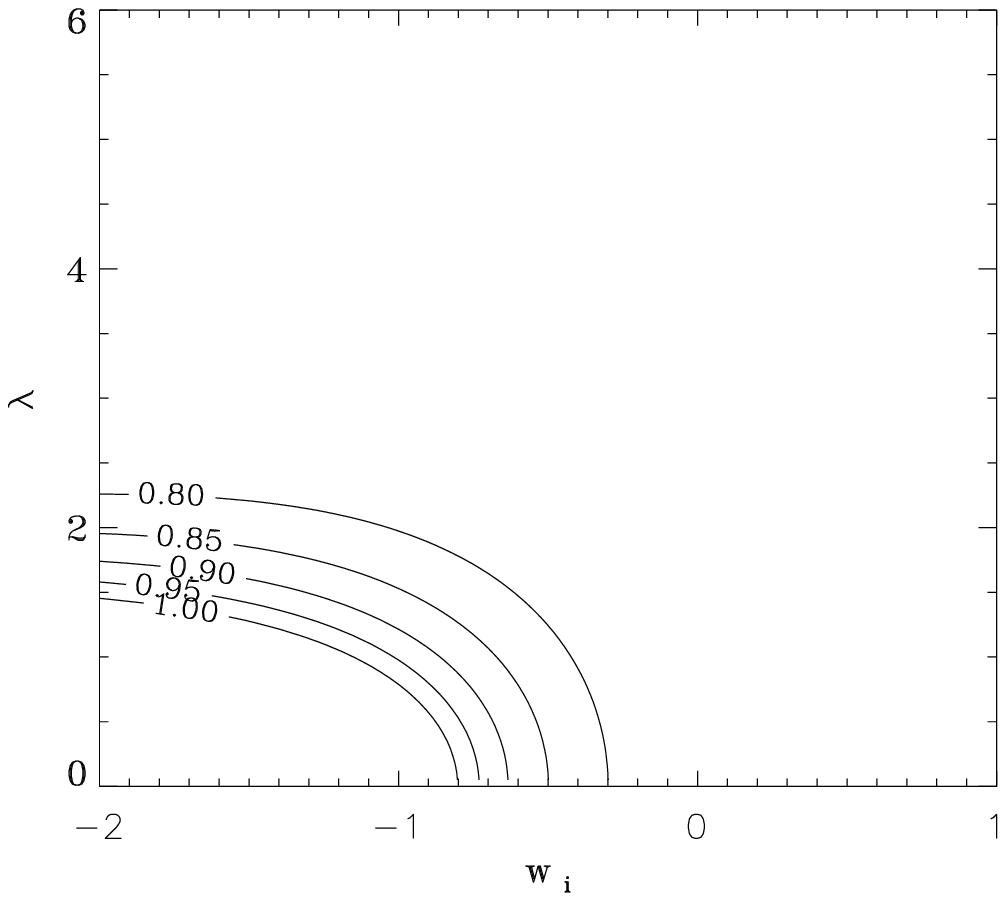}
\caption{\label{expV-H0t0} Contours of constant $H_0t_0$ in the $w_i$ -- $\la$ 
plane for the single exponential potential model.} \end{figure}

The integration proceeds until a value $H/H_0=1$ is reached. In 
Figs.~\ref{expV-Omph0} and \ref{expV-H0t0}, the contours of $\Omp=1-v_0^3$ 
and $H_0t_0$ are displayed in the $w_i$ -- $\la$ parameter space. In view of 
recent estimates of the ages of globular clusters \cite{Krauss}, a mean value 
of 12.8 Gyr for the age of the universe is arrive at. With $h\simeq0.7$ this 
would require $H_0t_0\sim0.9$. A concordant value of $\Omp=0.67\pm0.05$ is 
given by several observational tests \cite{WCOS}. By comparing these two 
contour plots, we see that parameters in the region of small $\la$ are 
favoured. This region shall represent models where the solutions approach the
attractor, either the scalar field dominated solution ($\la^2<3\gm$) or the 
scaling solution ($\la^2>3\gm$), only recently. Note that a seperate bound of 
$\la^2>20$ has been obtained from nucleosynthesis by assuming that the late 
times attractor is the scaling solution, and the attractor is approched within 
a few expansion times of the end of inflation \cite{CLW,FJ}. 

Another useful quantity which can be determined from the numerical integration
is the deceleration parameter 
\be\qn={3\over2}v_0^3+{1\over2}u_0^2-1\ .\ee 
$\qn<0$ corresponds to a universe whose expansion is accelerating at the 
present epoch. 

\subsection{Double exponential potential}

With the double exponential potential model, we again use the variables $u$ 
and $v$ as defined in (\ref{uvw}), and redefine $w$ as
\be
w=\sqrt{2}\kp\ph+\ln{A}\ .
\ee
In terms of these variables the field equations become
\bea
u'&=&-3{H\over H_0}u+\sqrt{2}V_1e^{w}\exp\left(-e^{w}\right)\ , \\
v'&=&-{H\over H_0}v\ , \\
w'&=&\sqrt{2} u\ ,
\eea
where we have replaced $V_0$ with a dimensionless parameter 
$V_1={\kp^2V_0\over H_0^2}$, and the Hubble parameter is
\be
{H\over H_0}\equiv\left[v^3+{1\over6}u^2+{1\over3}V_1\exp\left(-e^{w}\right)\right]^{1\over2}\ .
\ee

\begin{figure}[ht] \includegraphics[height=7cm,width=8cm]{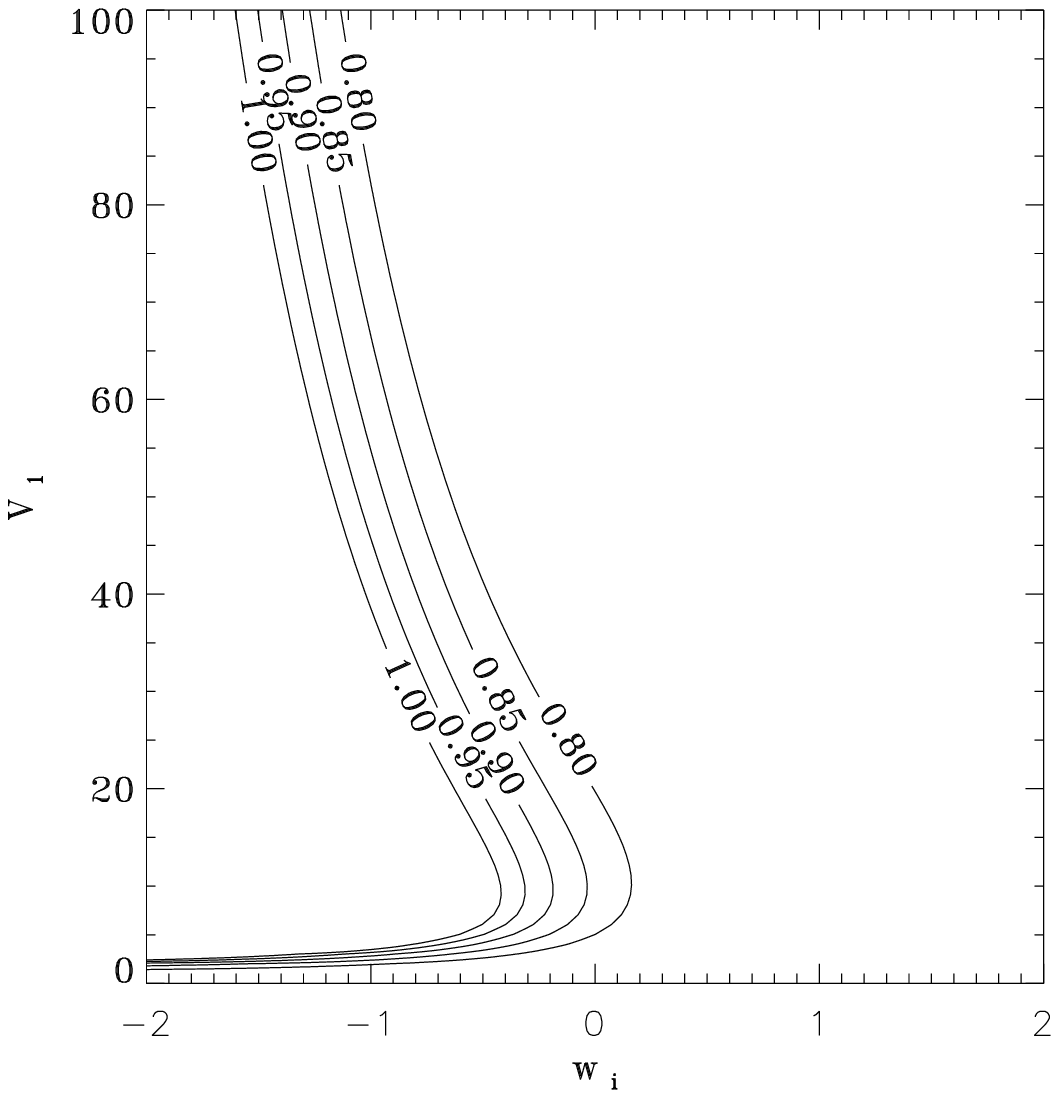} 
\caption{\label{dsb-H0t0} Contours of constant $\Omp$ in the $w_i$ -- $V_1$ 
plane for the double exponential potential model.} \end{figure}

\begin{figure}[ht] \includegraphics[height=7cm,width=8cm]{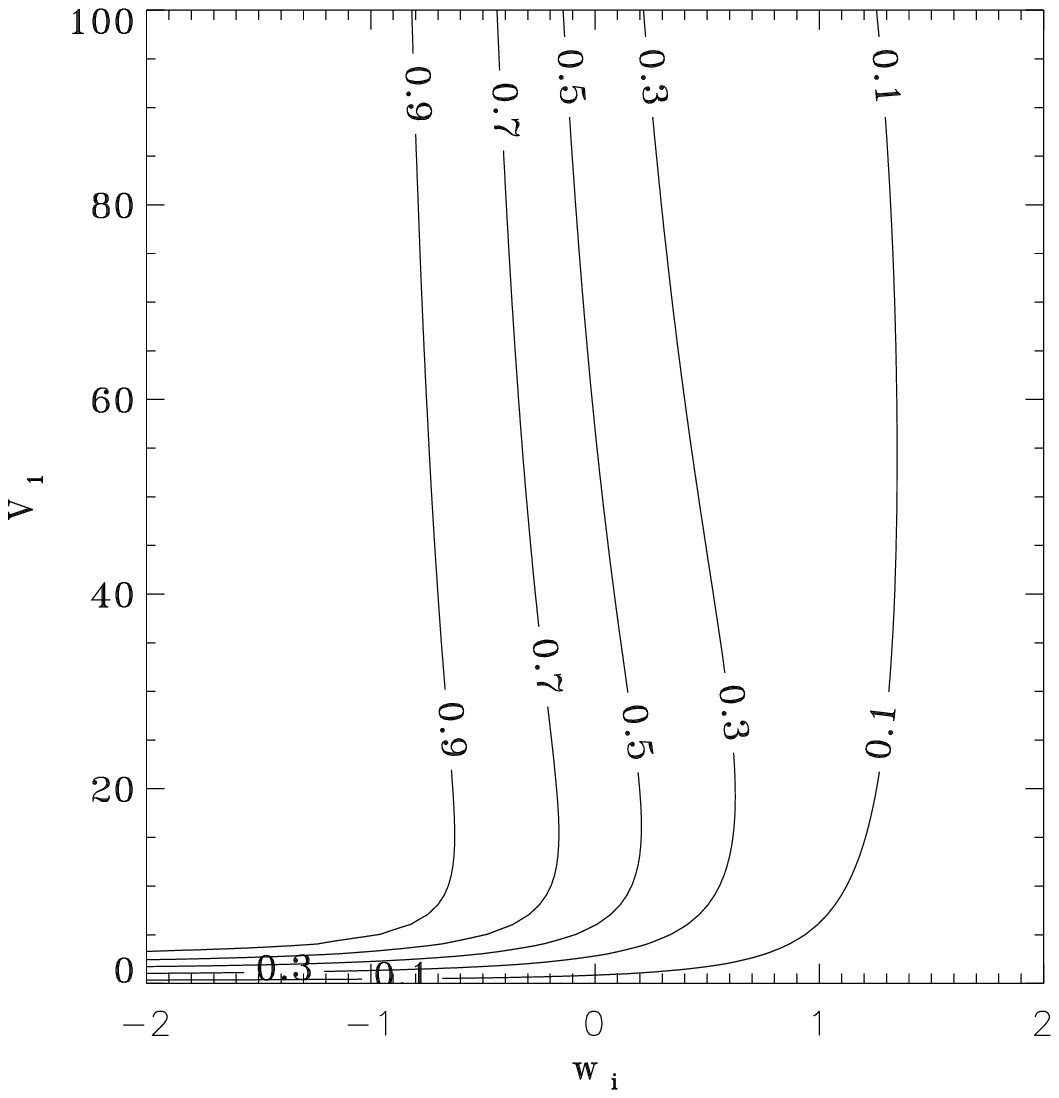} 
\caption{\label{dsb-Omph0} Contours of constant $H_0t_0$ in the $w_i$ -- $V_1$ 
plane for the double exponential potential model.} \end{figure}

We choose the same $H_0t_i$, $u_i$, and $v_i$ as above. The parameters 
$w_i$ and $V_1$ determine the phyisical origin of the model. The contours of 
$\Omp$ and $H_0t_0$ are displayed in the $w_i$ -- $V_1$ parameter space 
in Fig.~\ref{dsb-Omph0} and \ref{dsb-H0t0}. By comparing these two contour 
plots, the interesting region would seem to be the bottom left-hand corner of 
the parameter space. The interesting region corresponds to the scalar fields 
being still nearly frozen to their initial states, or having become dynamical 
and starting to roll down the potential slope only recently.

\section{Observational Constraints}

A constraint on the single exponential potential model has been obtained by
Frieman and Waga \cite{FW} using five type IA supernovae (Sne IA). In this 
section, we will constrain both the single and double exponential potential 
models using a larger set of Sne IA, together with a separate constaint using 
the gravitational lensing statistics of high luminosity quasars.

Empirical calibration of the light curve -- luminosity relationship of Sne IA 
provides absolute magnitudes that can be used as distance indicators. The 
Supernova Cosmology Project leaded by Perlmutter \etal\ \cite{Per} and the 
High Redshift Supernovae Search Team leaded by Riess \etal\ \cite{Rie} are two 
different groups have been working extensively in searching for high redshift 
Sne IA. I will use the larger available data set of the two, namely the 60 
Sne IA published by Perlmutter \etal\ \cite{Per}. The luminosity distance -- 
redshift relation so obtained provides a good observational constraint on 
quintessence models provided that there is no intrinsic evolution of the peak 
luminosities of the Sn IA sources. The issue of how the constraints are 
altered in the presence of such evolution has been discussed in \cite{Dre,NW}.

Gravitational lensing of distant quasars due to galaxies along the line of 
sight provides another relatively sensitive constraint \cite{Koc} on the 
quintessence models. The statistics of abundances of multiply imaged quasars 
and observed separations of the images to the source can be used to estimate 
the distances to the quasars. We follow the calculation as described by Waga 
and Miceli \cite{WM}, who used a total of 862 ($z>1$) high luminosity quasars 
plus 5 lenses from seven major optical survey.

\begin{figure}[ht] \includegraphics[height=8.5cm,width=8.5cm]{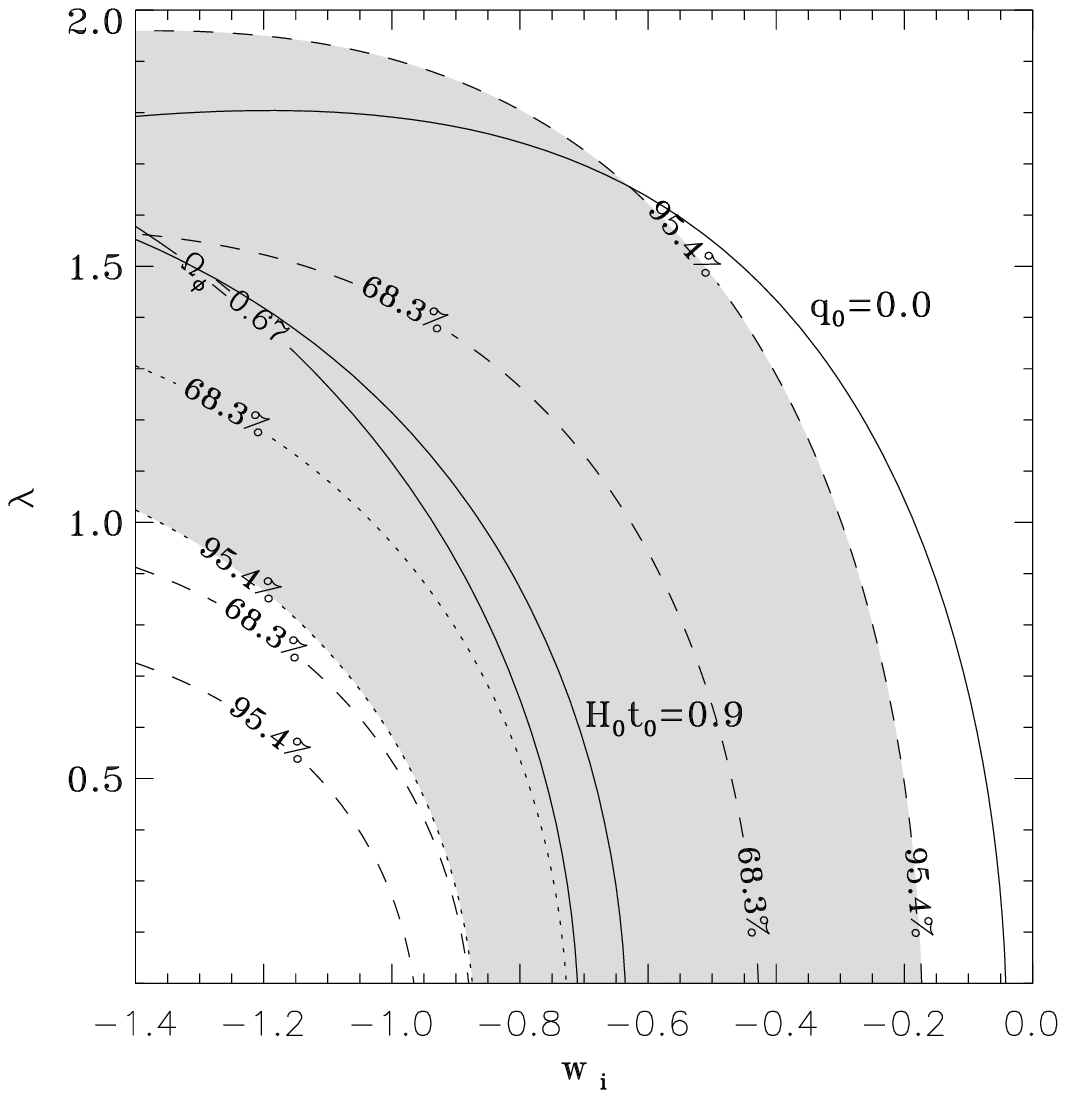}
\caption{\label{expV-oc} The 68.3\% and 95.4\% joint credible regions on the 
$w_i$ - $\la$ parameter space of the single exponential potential model, for 
the Sne IA test (dashed) and the gravitational lensing statistics test 
(dotted). Over plots are the contours of $H_0t_0=0.9$, $\Omp=0.67$, and  
$\qn=0$. The region allowed at 95.4\% by both observational tests is shaded.} 
\end{figure}

\begin{figure}[ht] \includegraphics[height=8.5cm,width=8.5cm]{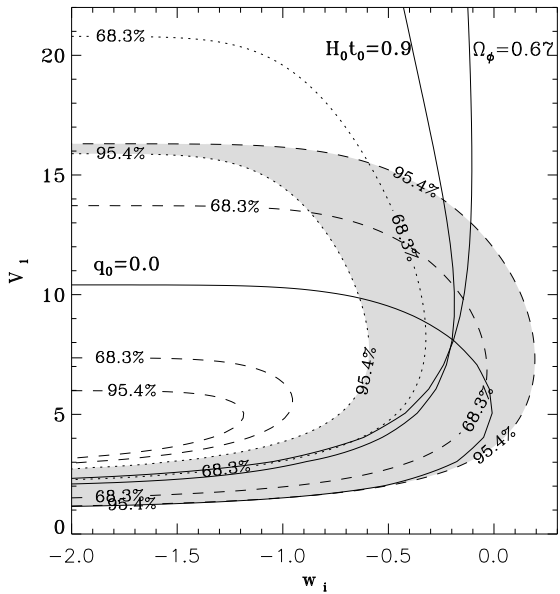}
\caption{\label{dsb-oc} The 68.3\% and 95.4\% joint credible regions on the 
$w_i$ -- $V_1$ parameter space of the double exponential potential model, for 
the Sne IA test (dashed) and the gravitational lensing statistics test
(dotted). Over plots are the contours of $H_0t_0=0.9$, $\Omp=0.67$, and 
$\qn=0$. The region allowed at 95.4\% by both observational tests is shaded.} 
\end{figure}

We display the results of both tests as 68.3\% and 95.4\% joint credible 
regions of the parameter spaces described in the previous section. 
Figs.~\ref{expV-oc} and \ref{dsb-oc} display results for the parameter spaces 
of the single and double exponential potential models respectively. For both 
models, the interesting regions that give rise to expected values of 
$\Omp\goesas0.67$ and $H_0t_0\goesas0.9$ are well within the 68.3\% confident 
region of both the Sne IA and the gravitational lensing statistics tests. The 
regions to the left of the $\qn=0$ contours correspond to models that give 
rise to accelerated expansion. 

\section{Discussion}

We have performed a phase-space analysis on the double exponential potential 
model, its properties can be understood in terms of a single exponential 
potential model with a varying coefficient $\la$. The analysis shows that 
unlike the single exponential potential model which has two late times 
attractor depending on $\la$, the double exponential potential model has only 
one global attractor which leads to a matter dominated universe at late times.
However, it is always possible for the model to be dominated by the scalar 
field at present and dominated by matter in the future. The problem with this 
model is we need to fine-tune the parameters in order not to have reached the 
attractor at the present epoch, thereby circumventing the objections of 
Binetruy \cite{Bin} and de. la. Macorra \cite{delaM}. However, all other 
quintessence models appear to also require a degree of fine-tuning 
\cite{PNGB,plp,ss,CLW,FJ,LS}.

We studied the observation constraints for both the single and double 
exponential potential models using a more updated type IA supernovae data and 
the gravitational lensing statistics. The results show that there are regions 
in the parameter spaces for which the models are consistent with the 
observations at the same time giving appropriate values for the scalar field 
energy density and the age of the universe at present. 

%%%%%%%%%%%%%%%%%%%%%%%%%%%%%%%%%%%%%%%%%%%%%%%%%%%%%%%%%%%%%%%%%%%%%%%%
\section*{Acknowledgments}

I would like to thank Chris Kochanek for supplying me with the gravitational
lensing data, Nelson Nunes, Ioav Waga, and David Wiltshire for helpful 
discussions about aspects of the paper.

%%%%%%%%%%%%%%%%%%%%%%%%%%%%%%%%%%%%%%%%%%%%%%%%%%%%%%%%%%%%%%%%%%%%%%%%
\def\PRL#1{Phys.\ Rev.\ Lett.\ {\bf#1}} \def\PR#1{Phys.\ Rev.\ {\bf#1}}
\def\ApJ#1{Astrophys.\ J.\ {\bf#1}} \def\AsJ#1{Astron.\ J.\ {\bf#1}}
\def\CQG#1{Class.\ Quantum Grav.\ {\bf#1}} \def\Nat#1{Nature {\bf#1}}
\def\JMP#1{J.\ Math.\ Phys.\ {\bf#1}} \def\NP#1{Nucl.\ Phys.\ {\bf#1}} 
\def\PL#1{Phys.\ Lett.\ {\bf#1}}

\end{document}